\documentclass{aa}  

\usepackage{graphicx}
\usepackage{txfonts}

\usepackage{color}

\newcommand{\chandra}{{\it Chandra} }

\newcommand{\hst}{{\it HST} }

\newcommand{\ergs}{${\rm erg \ cm^{-2} \ s^{-1}}$ }
\newcommand{\erg}{${\rm erg \ s^{-1}}$ }

\newcommand {\um}{$\mu$m~}
\newcommand {\ums}{$\mu$m}

\def\ts     {\thinspace}
\def\uJy    {$\mu$Jy}
\def\kms    {\ifmmode{{\rm \ts km\ts s}^{-1}}\else{\ts km\ts s$^{-1}$}\fi}
\def\msol   {\ifmmode{{\rm M}_{\odot}}\else{M$_{\odot}$}\fi}
\def\lsol   {\ifmmode{{\rm L}_{\odot}}\else{L$_{\odot}$}\fi}
\def\zsol   {\ifmmode{{\rm Z}_{\odot}}\else{Z$_{\odot}$}\fi}

\def\ltsima{$\; \buildrel < \over \sim \;$}
\def\simlt{\lower.5ex\hbox{\ltsima}}
\def\gtsima{$\; \buildrel > \over \sim \;$}
\def\simgt{\lower.5ex\hbox{\gtsima}}

\newcommand{\msun}{{\rm\,M$_\odot$}}
\newcommand{\sfr}{{\rm\,M$_\odot$\,yr$^{-1}$}}
\newcommand{\lsun}{{\rm\,L$_\odot$}}

\newcommand{\herschel}{{\it Herschel~}}
\newcommand{\src}{{\rm\,GH500.30~}}
\newcommand{\srcs}{{\rm\,GH500.30}}
\newcommand{\spitzer}{\textit{Spitzer}}

\begin{document}

   \title{A unique distant submillimeter galaxy with an X-ray-obscured radio-luminous active galactic nucleus 
   }


   \author{X. W.~Shu
          \inst{1}
          \and
          Y. Q.~Xue\inst{2, 3}
          \and
          D. Z.~Liu\inst{4}
          \and
          T. Wang\inst{6, 5}
          \and
          Y. K.~Han\inst{7}
          \and
          Y. Y.~Chang\inst{8}
          \and
          T.~Liu\inst{9}
          \and
          X. X.~Huang\inst{2}
          \and
          J. X.~Wang\inst{2, 3}
          \and
          X. Z.~Zheng\inst{10}
          \and 
          E.~da Cunha\inst{11}
          \and
          E.~Daddi\inst{5}
          \and
          D.~Elbaz\inst{5}
          }

   \institute{
   Department of Physics, Anhui Normal University, Wuhu, Anhui, 241000, China; \\
              \email{xwshu@mail.ahnu.edu.cn} 
            \and 
             CAS Key Laboratory for Researches in Galaxies and Cosmology, Department of Astronomy, University of Science and Technology of China, Hefei, Anhui 230026, China
             \and
             School of Astronomy and Space Science, University of Science and Technology of China, Hefei 230026, China
             \and
             Max Planck Institute for Astronomy,  K\"onigstuhl 17,  D-69117 Heidelberg, Germany
              \and
              Laboratoire AIM-Paris-Saclay, CEA/DSM/Irfu - CNRS - Universit\'e Paris Diderot, CEA-Saclay, F-91191 Gif-sur-Yvette, France 
              \and
              Institute of Astronomy, The University of Tokyo, 2-21-1 Osawa, Mitaka, Tokyo 181-0015, Japan
              \and
              Yunnan Observatories, Chinese Academy of Sciences, Kunming, 650011, China 
              \and
              Academia Sinica Institute of Astronomy and Astrophysics, PO Box 23-141, Taipei 10617, Taiwan
              \and
              Max-Planck-Institut f\"ur extraterrestrische Physik, Giessenbachstrasse 1, D-85748 Garching 
              bei M\"unchen, Germany
              \and
              Purple Mountain Observatory, Chinese Academy of Sciences,
              Nanjing, 210008, China
              \and
              The Australian National University, Mt Stromlo Observatory, Cotter Rd, Weston Creek, ACT 2611, Australia
             }


 
  \abstract
   {} 
   {
   We present a multiwavelength study of an atypical submillimeter galaxy, \srcs, in the GOODS-North field, 
   with the aim to understand its physical properties of stellar and dust emission, as well as the central active galactic nucleus (AGN) activity. 
   Although it is shown that the source is likely an extremely dusty galaxy at high redshift, 
   its exact position of submillimeter emission is unknown. 
}
   { 
 {We use NOEMA observation at 1.2 mm with subarcsecond resolution to resolve the dust emission, and
     precisely localize the counterparts at other wavelengths, which allows us to better constrain its
       stellar and dust spectral energy distribution (SED) as well as redshift.
  } We carry out the new near-infrared (NIR) photometry of \src observed with HST, and 
     perform panchromatic SED modelling from ultraviolet (UV)/optical to submillimeter. 
     We derive the photometric redshift using both NIR and far-infrared (FIR) SED modeling, 
     and place constraints on the stellar and dust properties such as stellar mass, age, dust attenuation, IR luminosity, 
     and star-formation rate (SFR). The AGN properties are inferred from the X-ray spectral analysis and 
     radio observations, and its contribution to the total IR luminosity is estimated from the broadband 
     SED fittings using {\tt MAGPHYS}. 
    } 
   {
   With the new NOEMA interferometric imaging, 
   we confirm that the source is a unique dusty galaxy. 
   It has no obvious counterpart in the optical and even NIR images 
   observed with \hst at $\lambda\simlt1.4$\ums. Photometric-redshift analyses from both stellar and dust SED suggest 
   it to likely be at $z\simgt4$, though a lower redshift at $z\simgt3.1$ cannot be fully ruled out (at 
   90\% confidence interval). 
 Explaining its unusual optical-to-NIR properties requires an old stellar population ($\sim0.67$ Gyr), coexisting 
 with a very dusty ongoing starburst component. 
 The latter is contributing to the FIR emission, with its rest-frame UV and optical 
 light being largely obscured along our line of sight. 
 If the observed fluxes at the rest-frame optical/NIR wavelengths were mainly contributed by old stars, a total 
 stellar mass of $\sim3.5\times10^{11}$\msun~would be obtained. 
 An X-ray spectral analysis suggests that this galaxy harbors a heavily obscured AGN {with
 $N_H=3.3^{+2.0}_{-1.7}\times10^{23}$ cm$^{-2}$} and an intrinsic 2-10 keV luminosity of $L_X\sim2.6\times10^{44}$ \erg, 
 which places this object among distant type 2 quasars.
 The radio emission of the source is extremely bright, which is an order of magnitude higher than the
 star-formation-powered emission, making it one of the most distant radio-luminous dusty galaxies.
  }
{The combined characteristics of the galaxy suggest that 
the source appears to have been caught in a rare but critical transition stage in the
evolution of submillimeter galaxies, where we are witnessing the birth of a young AGN and 
possibly the earliest stage of its jet formation and feedback.}

\titlerunning{A Distant Submillimeter Galaxy with An X-ray Obscured Radio-luminous AGN}
\authorrunning{X. W. Shu et al.}
 \maketitle
%
\section{Introduction}

Over the last decade, far-infrared (FIR) and submillimeter (submm) surveys have revolutionized
our understanding of galaxy formation and evolution 
by uncovering a new population of heavily dust-obscured starburst 
systems at high redshifts \citep[e.g.,][]{chapman05, michalowski10, casey14}.
These dusty submillimeter galaxies (SMGs) are found to have very high FIR
luminosities of $L_{\rm FIR}\sim10^{12}-10^{13}$\lsun~, 
indicative of extreme star formation rates (SFRs) of hundreds to thousands of solar masses per year. 
While the bulk of this population has been found to lie at $z\sim2$, 
recent blank-field surveys (e.g., by \herschel and South Pole Telescope) have started to 
reveal the high-redshift ($z\simgt4$) tail of the SMG distribution 
\citep[][]{Vieira13, dowell14, ivison16}.
These are thought to be progenitors of present-day massive ellipticals \citep[e.g.,][]{swinbank06}, 
and provide critical physical conditions for 
the formation of luminous quasars at high redshifts \citep[][]{coppin08}. 
However, details about their evolutionary process are still poorly known.  



Submillimeter galaxies are known to show red optical to near-infrared (NIR) color and
to not emit strongly in the rest-frame ultraviolet (UV) \citep[e.g.,][]{yun08}, 
mainly due to the huge dust extinction produced by ongoing starbursts 
and/or evolved stellar population if at high redshift.   
 Two illustrative examples are given by the bright SMGs GN10 and HDF850.1 in 
the Great Observatories Origins Deep Survey (GOODS) North field, both of which 
have no detectable optical or NIR counterparts in extremely deep imaging  
\citep[][]{wang07, cowie09}, 
and now have been confirmed to be at $z=4.05$ \citep[][]{daddi09a} 
and $z=5.2$ \citep[][]{walter12}, respectively, based on molecular CO emission lines. 
The two galaxies are instead detected significantly at mid-IR (MIR) wavelengths
$\lambda\simgt3.6$\um with the \spitzer/IRAC, yielding extremely red colors
of $H$-[4.5\ums]$>$4.0. 

An outstanding issue relevant to the dusty starburst galaxies 
at high redshift is the early growth of the central supermassive black holes (SMBHs). 
The volume density and redshift distribution of SMGs similar 
to those of quasars suggest that the two populations are possibly evolutionarily linked 
\citep[][]{chapman05}.  
Multiwavelength diagnostics, particularly those at X-rays, indicate
that a substantial fraction (20\%--50\%) of bright SMGs 
have an active galactic nucleus (AGN)
at their center 
\citep[][]{alexander03a, alexander05, Borys05, laird10, wang13}. 
Follow-up studies have found that the BH masses for SMGs
appear to be systematically smaller than those found in quasars 
\citep[][]{alexander08b}, implying that they are likely at an early 
phase of the BH growth.  

While the detections and identifications of SMGs at $z\simgt4$ are rapidly increasing 
\citep[e.g.,][]{capak08, coppin08, daddi09b, dowell14, ivison16}, 
the incidence and properties of AGNs (often heavily obscured) in
these high-redshift systems have yet to be properly explored. 
Only a few $z\simgt4$ SMGs with an X-ray AGN have been identified, in which the 
X-ray data are usually too poor to ensure meaningful spectral fitting and hence  
effectively constrain the amount of gas obscuring the nucleus \citep[e.g.,][]{wang13}.  
One notable example is the V-band dropout galaxy LESS J033229.4-275619 at z = 4.76 
\citep[][]{vanzella06}. 
This object was detected as a bright SMG \citep[][]{coppin09}, 
and the 4Ms $Chandra$ Deep Field-South observations \citep[][]{xue11} revealed the presence of a Compton-thick AGN  
with an X-ray luminosity of $L_{\rm X}\sim2.5\times10^{44}$\erg 
\citep[][]{gilli11, gilli14}, 
making it one of the most distant type 2 quasars discovered so far. 
Objects like LESS J033229.4-275619 are therefore ideal
laboratories to study the co-evolution of BHs with their host galaxies.



In this {paper}, we describe the multiwavelength properties of an unusual   
FIR bright galaxy, \src (RA=12:37:13.86; DEC=62:18:26.2),  which is undetected in most optical and 
even NIR bands with HST in the GOODS-North field, likely an extremely 
dusty galaxy at high redshift of $z\simgt4$. 
The source is not only extreme for its faintness from optical to NIR, but the X-ray and radio 
properties are exceptional. For instance, 
it is extremely bright in the radio for a SMG, with a 1.4 GHz flux density of $\sim630\mu$Jy, 
and has a radio excess of an order of magnitude more than the star-formation powered emission, 
indicating that it maybe one of the most distant radio-luminous SMGs in the early Universe. 
We note that some results from X-ray observations of the source  
have been presented by \citet[][]{laird10} for a sample of bright SMGs.  
However, \citet[][]{laird10} did not perform a detailed analysis of the 
multiwavelength spectral energy distribution (SED), and in particular 
quoted a rough estimate of photometric redshift ($z=2.6$) from only IRAC colors, 
{which may be incorrect and may therefore bias our understanding of the physical properties of the source.}
In addition, their reference for the high submillimeter flux at 850\um is 
found to be incorrect in later observations (see Section 2). 
We will provide a revised view on the true nature of this unique
SMG, in particular in light of the new NOEMA data which allow for 
precise localization of submillimeter emission and hence unambiguous 
identification of its counterparts at other bands. 
Throughout this paper,
we assume a cosmology with $H_0$ = 70 km s$^{-1}$ Mpc$^{-1}$, $\Omega_M$ = 0.27, $\Omega_\Lambda$ = 0.73.

\section{Data and analysis}
\label{sec:data}

\subsection{Optical to mid-infrared data}

The optical to MIR data we used for photometric analysis are drawn from a $K_S$-band 
selected multiwavelength 
catalog ($K_s=24.3$ mag, ID=73108, Yang et al. 2014) supplemented 
by the new \hst data from the 3D-\hst program \citep[][]{skelton14}. 
{The 3D-\hst data release\footnote{3dhst.research.yale.edu/Data.php} provides 
calibrated ground-based and HST images, as well as multiwavelength photometric catalogs for five CANDELS fields. 
The photometric catalog for each field was constructed based on the sources detected 
in the combined image of three \hst/WFC3 NIR bands (F125W, F140W, and F160W). 
Matching with the coordinate of the Ks-band source for GH500.30 (12:37:13.86; +62:18:26.43), we did not find 
a counterpart 
in the 3D-\hst photometric catalog suggesting non-detections in the three WFC3/IR bands (F125W, F140W and F160W) down to 
a 5$\sigma$ depth of $\sim$26.2 mag. We attempted to perform manual photometry at the position of the $K_S$-band source 
in a circular aperture of radius 0.6\arcsec~ using the Python photometry routine developed
by ourselves to determine whether there
is any low-level flux recorded.  
The background and flux errors were estimated in an annulus
area with inner and outer radii of 1.5\arcsec and 3\arcsec, respectively.   
We found only a marginal detection of 26.6$\pm0.3$ mag in the $H$-band ($\sim$1.6\ums), and the source is not 
detected ($<1\sigma$) in any other \hst optical and NIR wavelengths (i.e., F435W, F606W, F775W, F850LP, F125W, and 
F140W). 
We provide the \hst measurements (and 1$\sigma$ upper limits) in Table 1. }


\subsection{FIR and submillimeter data}

We take advantage of the FIR data from deep {\it Herschel} observations as a part of the 
GOODS-{\it Herschel} program. A detailed description of the observations
is given in \citet[][]{elbaz11}.
{Since \src is well separated in the MIPS 24\um, PACS 100\um, and PACS\textbf{} 160\um images, the photometry
at these passbands is directly taken from the GOODS-Herschel catalog. 
Source confusion is a severe issue in the \herschel/SPIRE data where the spatial 
resolution is poor ($\theta_{\rm beam}\sim$20-40\arcsec), which makes it difficult to 
identify the correct counterparts and measure fluxes for individual Herschel sources. 
As we demonstrated in Shu et al. (2016), some candidate Herschel sources at $z\simgt2$ 
have no measured 500\um fluxes in the catalog of the GOODS-Herschel data release, 
which is likely due to the effect of source blending. 
This is for the case of \srcs, for which the non-detection in the 500\um in the catalog is likely due to 
the incorrect association of its flux with the neighboring galaxy (RA=12:37:16.83; DEC=62:18:24.16) at $z=0.84$.  
Therefore, the \herschel/SPIRE flux densities for the source were derived independently from our PSF-fittings 
with GALFIT (Peng et al. 2002). 
We used a very similar source extraction and photometry method to the one described in Elbaz et al. (2011),
but placed constraints on the 500\um flux density of the nearby source to be less than the 
predicted one from SED fittings. 
We fitted the 24, 100, and 160\um photometry with a library of CE01 templates 
to obtain a best-fit SED, using which we predicted the 500\um flux of $\sim$5 mJy 
for the nearby contaminating source.  
The \herschel/SPIRE fluxes from our GALFIT fittings are listed in Table 1. 
We note that our measurements are consistent with the super-deblended fluxes  
in the photometry catalog of Liu et al. (2018, ID 3532) where
$S_{250\mu m}=8.5\pm0.4$ mJy, $S_{350\mu m}=11.2\pm0.8$ mJy,  and $S_{500\mu m}=10.7\pm1.0$ mJy 
are found. 
Although there are a few nearby sources within a radius of $\sim$6\arcsec~ seen in the NIR images 
{(Fig. 1)} 
that might contribute to the 500\um flux,
our NOEMA observation at subarcsecond resolution (see below) revealed no other sources close to 
\src (with a 3$\sigma$ upper limit on 1.2 mm flux of $\sim$0.5 mJy, Fig. 1), indicating that 
the flux contamination from these neighboring sources on the FIR fluxes is negligible. 
}
\begin{table}
\caption{\scshape Photometric Data of \src.}
\centering
\begin{tabular}{llccl}
\hline\hline 
Telescope & Band  & Flux & Unit  & Ref. \\
\hline
KPNO &  3593\AA  & $<0.024$  & $\mu$Jy   & (1) \\
Subaru &  4458\AA   & $<0.018$  & $\mu$Jy & (1) \\
Subaru &  5478\AA   & $<0.041$  & $\mu$Jy  & (1) \\
Subaru &  6289\AA  & $<0.036$  & $\mu$Jy  & (1) \\
Subaru &  7684\AA  & $0.064\pm0.047$  & $\mu$Jy  & (1) \\
Subaru &  9037\AA & $<0.052$  & $\mu$Jy  & (1)\\
UH &  1.25\um  & $<0.073$  & $\mu$Jy & (1) \\
UH &  1.64\um & $<0.244$  & $\mu$Jy & (1)  \\
\hst & 4318\AA  & $<0.039$ & $\mu$Jy & (2) \\
\hst & 5919\AA & $<0.026$ & $\mu$Jy & (2)  \\
\hst & 7693\AA & $<0.052$ & $\mu$Jy & (2) \\
\hst & 9036\AA & $<0.077$ & $\mu$Jy & (2) \\
\hst & 1.247\um  & $<0.023$ & $\mu$Jy & (2) \\
\hst & 1.392\um  & $<0.040$ & $\mu$Jy & (2)  \\
\hst & 1.539\um  & $0.082\pm0.023$ & $\mu$Jy & (2) \\   
CFHT &  2.15\um & $0.723\pm0.15$  & $\mu$Jy & (1) \\
$Spitzer$  &  3.6\um   & $3.458\pm0.156$  & $\mu$Jy & (1)  \\
$Spitzer$  &  4.5\um  & $6.031\pm0.233$  & $\mu$Jy & (1) \\
$Spitzer$  &  5.8\um   & $10.07\pm1.048$  & $\mu$Jy & (1) \\
$Spitzer$  &  8.0\um   & $19.52\pm1.168$  & $\mu$Jy & (1)  \\
$Spitzer$ &  24\um   & $63.36\pm7.06$  & $\mu$Jy & (3) \\
\herschel &  100\um  & $1.08\pm0.35$  & mJy & (3) \\
\herschel &  160\um  & $3.45\pm0.95$  & mJy & (3) \\
\herschel &  250\um   & $10.63\pm2.02$  & mJy & (2) \\
\herschel &  350\um  & $12.45\pm2.76$  & mJy & (2) \\ 
\herschel&  500\um  & $11.5\pm3.19$  & mJy  & (2) \\
JCMT &  850\um  & $4.49\pm0.82$  & mJy  & (4) \\
JCMT &  1100\um  & $2.39\pm1.1$  & mJy  & (5) \\
NOEMA & 1200\um          & $2.24\pm0.16$ & mJy  & (2) \\
{VLA} &  1.4 GHz   & $626\pm5$  & $\mu$Jy  & (4) \\
{VLA} &  5.5 GHz  & $270.1\pm7.1$ & $\mu$Jy  & (6) \\
{VLBI} & 1.6 GHz  & $252.5\pm10.1$ & $\mu$Jy  & (7) \\ 
\chandra &  0.5--2 keV   & $4.8\times10^{-16}$  & \ergs & (2) \\
\chandra &  2--10 keV   & $1.5\times10^{-15}$  & \ergs & (2) \\
\hline
\end{tabular}
\tablefoot{(1) Yang et al. 2014; (2) This work; (3) {Elbaz et al. (2011)}; (4) Barger et al. 2014; 
(5) {Perera et al. 2008; (6) Guidetti et al. 2017; (7) Radcliffe et al. 2016}.
}
\end{table}

The source was also detected in the SCUBA2 850\um with $S_{850\mu m}=4.5\pm0.8$ mJy (CDFN28, Barger et al. 2014), 
and AzTEC 1.1mm with $S_{\rm 1.1 mm}=2.4\pm1.1$ mJy ({\rm AzGN26}, Perera et al. 2008) and is found to have similar 850\um flux in the recent SCUBA2 Cosmology Legacy Survey data release 
\citep[][]{geach17}. 
We note that the source was reported to have much larger 850\um flux 
($S_{850\mu m}=10.9\pm2.9$ mJy) based on the old SCUBA data \citep[][]{wall08}, 
which is likely attributed to the large contribution from noise artifact, 
as it lies at the edge of the map where the noise fluctuations are significant.   

{Recently, \src has been observed with NOEMA as a part of the program aiming at accurately 
measuring the FIR SED for a sample of \herschel selected galaxies at $z>3$ 
(Liu et al. in preparation).
The observations of \src were done in the winter of 2016 with 8 antennae in
track-sharing mode with other Herschel sources in the sample. 
The
on-source time for GH500.30 is $\sim$1 hour. The synthesized beam is
$\theta_{\rm maj}\times\theta_{\rm min}$$\sim$1.06\arcsec$\times$0.63\arcsec\  and the continuum of this source is well detected
(spatially unresolved) at the phase center which is predefined as its
IRAC position.
With the NOEMA data, we obtained an accurate position for the submillimeter source (12:37:13.879; +62:18:26.296) 
and flux at 1.2mm ($S_{1.2\rm mm}=2.24\pm0.16$) mJy through $uv$-plane point-source fittings 
with the GILDAS\footnote{{http://www.iram.fr/IRAMFR/GILDAS}} uv\_fit tool.  
The statistical positional error can be estimated as  $\Delta\theta_{\rm stat}\simeq
\sqrt{\theta_{\rm maj}\theta_{\rm min}}/(2S/N)$ 
(where $\theta_{\rm maj}$ and $\theta_{\rm min}$ are the major and minor axes of the beam; 
e.g., Younger et al. 2007). 
Given the source's signal-to-noise ratio (S/N) of $\sim14$, we estimated the positional error  at $\Delta\theta_{\rm stat}\sim0.03$". 
Therefore, the high angular resolution and positional accuracy 
of the NOEMA data are good enough for us to precisely localize the source position and identify correct
counterparts at other bands without confusion problems. 
For example, the synthesized beam size of $\sim$1\arcsec\ at 1.2 mm is comparable with (or better than)  
the resolution at Ks-band (FWHM$\sim$0.8\arcsec) and IRAC (FWHM$\sim$2\arcsec). 
This confirms that the SMG has no optical counterparts in the deep \hst imaging. 

}


\begin{figure*}[th]
\centering
 \includegraphics[scale=0.95]{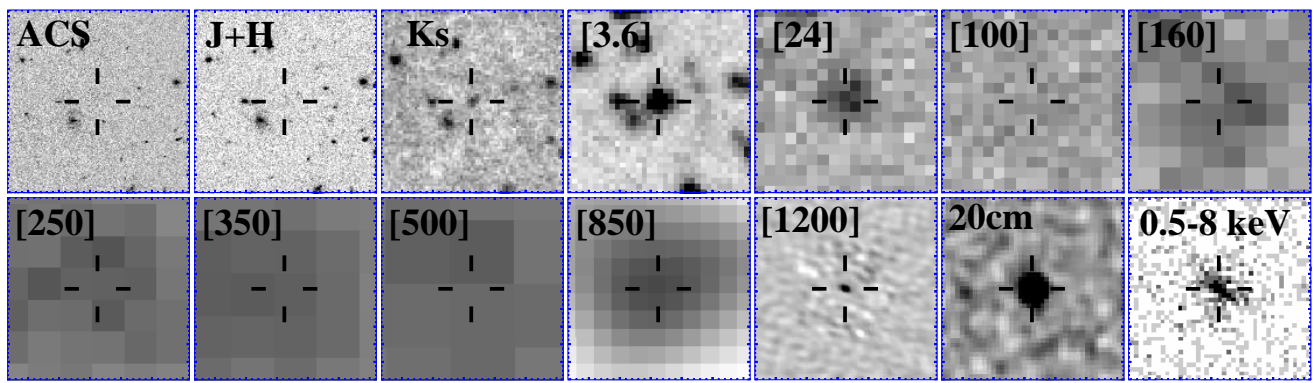}
\caption{
Postage-stamp images of \src from multiwavelength observations. From
left to right and top to bottom: 
HST/ACS (F435+F606+F775+F850LP), HST/WFC3 (F125W+F140W+F160W), CFHT $K_s$, IRAC [3.6\ums], 
MIPS 24\ums, PACS 100\um and 160\ums, SPIRE 250\ums, 350\um and 500\ums, 
SCUBA2 850\ums, NOEMA 1.2 mm, VLA 1.4 GHz, and \chandra 0.5-8
keV. Each panel has a size of 20\arcsec$\times$20\arcsec.  
The detailed photometry is listed in Table 1. 
}
\label{fig:image}
\end{figure*}


\subsection{X-ray data}
\label{2.2}
A total exposure of $\sim$2 Ms has been accumulated on the \chandra Deep Field North (CDFN) 
as a result of 20 individual observations with \chandra ACIS-I \citep[][]{alexander03b}, 
which is the second deepest X-ray survey ever (see, e.g., Xue 2017 for a review), behind the 7 Ms CDFS \citep[][]{luo17}. 
X-ray data products including event files for each observation and also for the merged 
data set are publicly available\footnote{http://cxc.harvard.edu/cda/Contrib/CDFN.html}. 
{The galaxy GH500.30} was clearly detected in the 2 Ms CDFN catalog (XID=374, Alexander et al. 2003b)
with the soft X-ray flux of $F_{\rm 0.5-2~keV} =4.71\times 10^{-16}$ \ergs 
and the hard X-ray flux of $F_{\rm 2-8~keV} = 1.19\times 10^{-15}$ \ergs. 
In this paper, we used the new CDFN products with the improved data reduction by 
\citet[][]{xue16}, which includes a flexible and reliable two-stage
source-detection approach, better background estimate, and optimal extractions of X-ray photometry.   
We note that the separation between the $K_S$-band and X-ray centroid is $\sim$0.27\arcsec, which is well within the 
1$\sigma$ X-ray source positional uncertainty ($\sim$0.47\arcsec). 
We obtained 235$\pm17$ net counts in the 0.3-7 keV band, corresponding to a $13.8\sigma$ detection.  
We checked the photometry of five periods separately and found no significant source variability in both soft and hard bands.
A detailed X-ray spectral analysis will be presented in Sect. 4.1.

\subsection{Radio data}
{The radio flux density at 1.4 GHz ($S_{1.4\rm GHz}=626\pm5\mu$Jy) was taken from 
Barger et al. (2014), which was also listed in the source catalog of the latest 
VLA observations of the GOODS-North field (Owen 2018). 
The latest VLA imaging reaches an rms noise in the field center of 2.2\uJy\ with a resolution of 1.6\arcsec\ .  
}
The source is also detected at 5.5 GHz with {VLA} at a sub-arcsec resolution, with 
a flux $S_{5.5\rm GHz}=270.1\pm7.1\mu$Jy \citep[][]{guidetti17}. 
A compact radio core emission at 1.6 GHz has been detected with the new VLBI observations at 
a resolution of $\sim$5 mas \citep[][]{radcliffe16}, which yields a flux of $S_{1.6\rm GHz}=252.5\pm10.1\mu$Jy. 
The radio position is coincident with the position found in the submillimeter imaging obtained at 
the NOEMA.

%


\section{SED and redshift constraints}

We used the position of the submillimeter source obtained with NOEMA to identify 
its counterparts at other bands; namely, X-ray, optical to MIR, FIR and radio. 
There are no obvious counterparts in the optical and even NIR observed with \hst at the position of the 
submillimeter source. 
The flux densities and/or upper limits from optical to radio wavelengths are given in Table 1. 
The postage-stamp images of \hst optical and NIR, CFHT $K_s$, \spitzer~IRAC and MIPS, 
\herschel~PACS and SPIRE, SCUBA2 850\ums, NOEMA 1.2mm, VLA 1.4 GHz, 
and $Chandra$ X-ray are presented in Fig.~\ref{fig:image}. 
The weak detections in the optical and NIR reveal an extraordinarily red SED in the NIR. 
For example, the marginal detection at the 1.6\um by \hst implies a $S_{4.5\mu m}/S_{1.6\mu m}$ flux 
ratio greater than 70, corresponding to an extremely red color of $H$-[4.5\ums]$>$4.6. 
We note that it is unlikely that this red color is due to the Lyman break which would 
otherwise imply a redshift $z>16$.  
Galaxies with red $H$ and IRAC colors are identified to be massive dusty galaxies at either 
$z\sim2.5$ or $z\sim4.4$ \citep[][]{wang16}, but most of them have $H$-[4.5\ums] colors in the range 
of $2<H-[4.5\mu m]<3$.  
 

It is extremely difficult to measure the redshift of \src through conventional optical 
and NIR spectroscopy because of its faintness in the optical and NIR. 
Here we estimate the photometric redshift (photo-z) by fitting the photometric data points 
with a suit of SED templates to gain further insights into its nature. 

\subsection{Photometric redshift from near-infrared data}

Since the \src has extremely red $H$-[4.5\um] color indicating unusually high dust extinction 
\citep[e.g.,][]{huang11}, 
the standard templates used by most photo-z codes may not be enough to 
represent the observed SED of the source.
We use Bayesian SED modeling code--BayeSED \citep[][]{han12, han14} to self-consistently estimate 
the photometric redshift and infer the physical properties of the galaxy. 
By employing efficient machine learning methods, such as the artificial neural network (ANN) 
algorithm, K-Nearest Neighbors (KNN) searching, and the Bayesian inference tool--MultiNest, 
BayeSED allows a continuous stochastic sampling of the parameter space and provides the posterior 
probability distribution function of all modeling parameters. This allows us to easily recognize 
the degeneracies between parameters and to  reasonably estimate the parameter uncertainties. 
We note that the BayeSED code has been extensively tested for both mock and real galaxy catalogs 
in \citet[][]{han14}, and found to show consistent results with the traditional 
grid-based SED-fitting code FAST \citep[][]{kriek09}. 
 

We have used Bruzual \& Charlot (2003, BC03) stellar population synthesis models 
with solar metallicity, assuming the \citet[][]{chabrier03} initial mass function (IMF) form 
in the range 0.1--100 \msun.  
We adopted an exponentially declining star-formation history (SFH) with $e$-folding times in the range of $\tau=0.1-10$ Gyr. 
For our photometric redshift calculations we adopted uniform priors on log(age/yr) in the range of 
[7.0, 10.1], and a very wide range of dust extinction of $0<A_V<15$ mag assuming the standard 
Calzetti attenuation curve.
As another prior, the age of a galaxy is restricted to
be less than the age of the universe at that redshift of the
galaxy. 

\begin{figure}[th]
\centering
\includegraphics[scale=0.5]{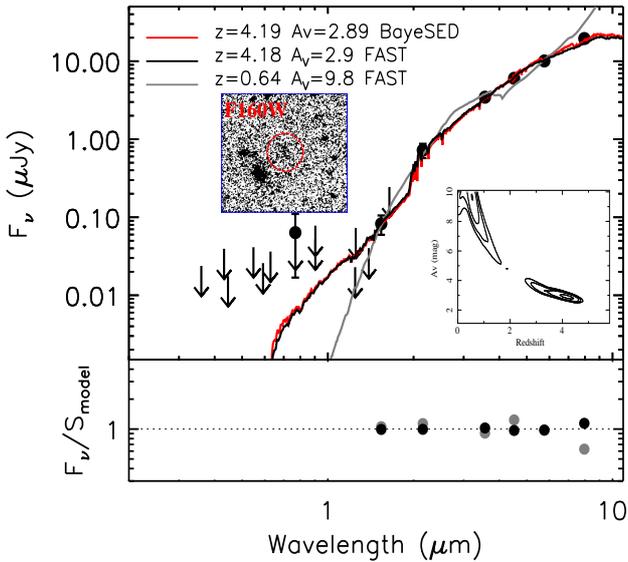}
 \caption{
Optical to MIR SED for \srcs {(upper panel)}. 
The filled circles show the observed photometry, while the downward arrows indicate $1\sigma$ 
upper limits.  
The red curve shows the best-fit (maximum likelihood) SED based on BayeSED modeling, 
whilst the black curve represents the best-fit SED from FAST.  
We also plot the fit with an extremely dusty template ($A_V=9.8$ and $z_{\rm phot}=0.64$) for comparison 
(gray curve).  
The \hst H-band image is shown in the inset left panel. 
Insert right panel shows the $\chi^2$ contours of redshift vs. dust extinction ($A_V$) 
from the FAST SED fittings.
{The lower panel shows the data to model ratios for the BayeSED fitting at z=4.19 (black solid circles) 
and for the FAST SED fitting at z=0.64 (gray solid circles). 
Only detected fluxes above 1\um are shown. }
}
\label{fig:nirsed}
\end{figure}

We fitted the model templates to the U-band to IRAC 8\um photometry (including 
1$\sigma$ upper limits) listed in Table 1 to investigate which redshift 
and how much dust extinction can make the red SED of the \srcs. 
Although the source is not detected in most optical and NIR bands, the upper limits suggest 
a very red SED at $\lambda<$2\um (Fig.~\ref{fig:nirsed}) and therefore still provide useful constraints. 
We assigned zero fluxes to the non-detections and corresponding $1\sigma$ errors as the 
flux uncertainties.   
The best-fit (maximum likelihood) template from BayeSED yields a redshift at $z=4.19\pm1.12$ and 
a modest extinction of $A_V=2.89\pm1.8$.
Such a model template is similar to those obtained in \citet[][]{huang11} for a sample 
of dusty galaxies with similarly red $H$-[4.5\ums] colors. 
We note that using the $EAZY$ photo-z package \citep[][]{brammer08} with empirically observed SEDs 
yields a consistent but slightly higher redshift at $z=4.56$ (1$\sigma$ confidence range 4.21--5.03).

{For a consistency check, we also used FAST to perform 
SED fittings to the U-band to IRAC 8\um photometry.  
We assumed the solar metallicity and 
exponentially declining SFH with $e$-folding times in the range $\tau=0.1-10$ Gyr. 
We adopted the standard Calzetti law with a very wide range of dust extinction of $0<A_V<10$. 
The best-fit template from FAST yields a redshift at $z=4.18$ with a modest extinction of $A_V=2.9$, 
which are in good agreement with the fitting results from the BayeSED. 
The best-fit SED templates are shown in Fig.~\ref{fig:nirsed}.  
}

\begin{figure}
\centering
 \includegraphics[scale=0.6]{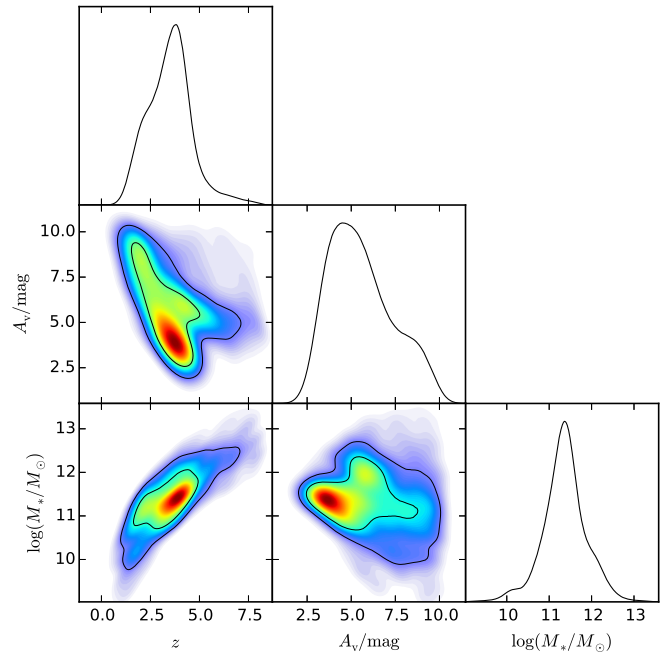}
\caption{
Joint one- and two-dimensional posterior probability distributions 
(normalized to unit area) of redshift, dust extinction $A_V$, and stellar mass for \srcs.  
The color coding represents confidence levels. The joint 90 and 95\% confidence contours 
are shown in black. 
}
\label{fig:contour}
\end{figure}

One uncertainty in the above photo-z analysis is the degeneracy between 
the redshift and dust extinction in the SED fittings. 
As we show in Fig.~\ref{fig:nirsed}, if we allow very large extinctions, the 
red SED of \src can be reproduced by a highly reddened galaxy ($A_V\sim10$) at $z<1$. 
However, we argue that this possibility of an extremely dusty low$-z$ galaxy can be 
ruled out given its significant FIR to submillimeter emission. 
Furthermore, SED fittings to submillimeter-selected galaxies generally give $A_V\sim0-5$ 
\citep[][]{yun12, dacunha15}. 
For example, the stellar population analysis for the GN 10, which is a known dusty galaxy at $z=4.05$, yields 
$A_{V}\sim3$ \citep[][]{daddi09a}. This value is very similar to that obtained from 
our best-fit (maximum likelihood) template for the \srcs. Figure~\ref{fig:contour} shows 
the posterior probability distributions on the redshift, dust extinction $A_{\rm V}$ and the stellar mass 
from the BayeSED fittings. 
Although degeneracies in stellar population modeling parameters do exist, the 
Bayesian analysis suggests that significant dust extinction (i.e., $A_{\rm V}>7$) is unlikely. 

Another uncertainty is the existence of an AGN, 
which may have effects on the NIR photometric analyses. 
The galaxy \src shows a featureless power-law continuum in the IRAC bands, with the brightest being the 
8\ums; if due to the stellar emission, this would imply that the rest-frame 1.6\um bump is beyond 
the 5.8\um channel, or $z\simgt3.5$. However, the IRAC emission can also be due to {a reddened and 
obscured AGN}, as indicated by a slight excess emission at 8\um compared with the 
best-fit SED template (Fig.~\ref{fig:nirsed}). \citet[][]{yun08} proposed that 
a young dusty starburst at high redshift is able to 
generate similar red IRAC colors as obscured AGNs. The galaxy \src appears to fall into the red power-law region 
in the IRAC color-color diagram of \citet[][]{yun08}, where AGNs and 
young and extremely dusty starbursts at $z>3$ overlap. 
{Although it is difficult to disentangle 
the relative contribution of the two causing the IRAC power-law spectral slope},  
{we compared the $S_{500\mu m}/S_{24\mu m}$ flux ratio to  Mrk
  231, a local FIR luminous Compton-thick AGN. 
For a fair comparison we redshifted the SED of Mrk 231 to z=4.19 (the best-fit photometric
redshift for \srcs) to derive the $S_{500\mu m}/S_{24\mu m}$ ratio. 
We found the $S_{500\mu m}/S_{24\mu m}$ flux ratio of \src is a factor of 3.9
higher than that in Mrk 231, indicating that the source may have low levels of the 
warm dust emission due to an AGN.} 
In fact, we estimated that an AGN may contribute at an approximately $12^{+23}_{-8}$\% level 
to the rest-frame MIR continuum emission (Sect. 4.3). 
We note that by using only a library of pure AGN- or hybrid AGN-galaxy templates\footnote{The templates were used to derive 
photometric redshifts for bright X-ray sources in the COSMOS field \citep[][]{salvato09}.}, we obtained a 
best-fit photometric redshift at $z_{\rm phot}=4.75^{+0.85}_{-0.49}$, a value consistent  
with that estimated assuming stellar templates within errors.
\citet[][]{yang14} obtained a similar best-fit redshift of $z_{\rm phot}=
4.43^{+1.64}_{-0.23}$ by using an optimal combination of galaxy and AGN templates. 
We therefore conclude that the existence of an AGN has a small effect on the photo$-z$ estimation of \srcs.
In fact, the AGN's emission at the rest-frame UV/optical is likely heavily
attenuated considering the strong absorption inferred from the X-ray observations (Sect. 4.1 ). 



\subsection{Photometric redshift from far-infrared SED} 

An independent estimate of the redshift for \src can be performed from its FIR emission, 
because the mid-to-far-IR emission has a generally well-defined peak at the rest-frame 50-200\ums. 
There have been a number of previous studies exploring the use of the FIR-radio relation to constrain the redshifts of SMGs 
\citep[e.g.,][]{carilli99, wang09, barger12}, 
which would be helpful when a galaxy suffers from extremely heavy extinction in the rest-frame optical/NIR bands.
The main uncertainty of this method is the dust temperature which is still poorly constrained for high-$z$ SMGs. 

\begin{figure}[]
\centering
\includegraphics[scale=0.8,angle=0]{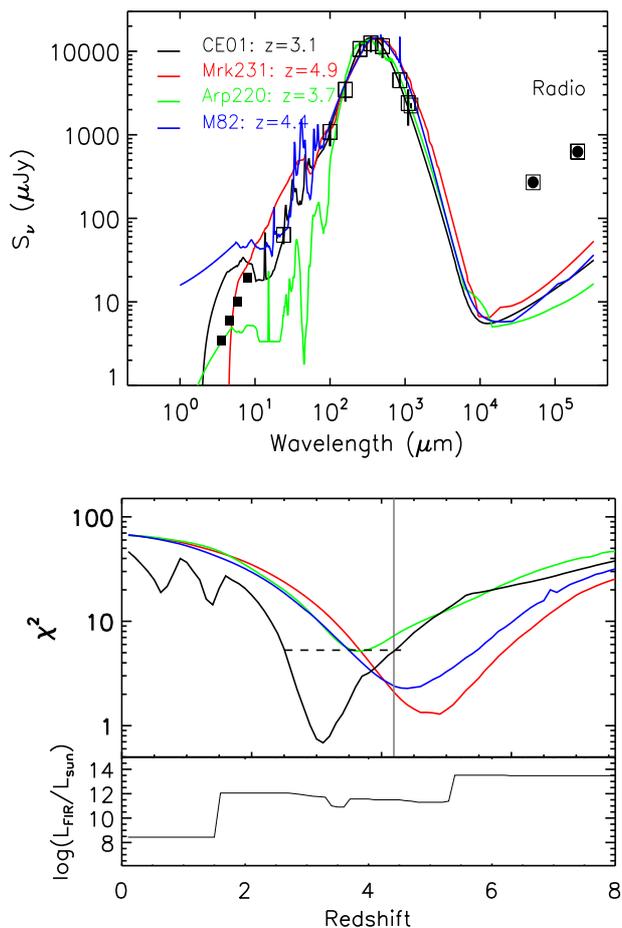}
 \caption{{\it Top panel}: 
FIR to radio SED of \src (open squares). The black curve shows the best-fit SED 
among all 105 CE01 templates. 
The best-fits with Mrk 231, Arp 220 and M82 SED are shown in red, green and blue curves, 
respectively. The fill squares are observed \spitzer/IRAC fluxes.  
{\it Bottom panel}:  $\chi^2$ vs. redshift. The black curve represents the results from the best-fit 
CE01 SED at each redshift, and the color-coded curves are for the Mrk 231, Arp 220 and M82 SED. 
The horizontal dashed line shows the 90\% confidence interval for the redshift. 
The lower panel at the bottom shows the IR luminosities from CE01 templates that allow for  
the minimum $\chi^2$ at each redshift.  
The vertical gray curve represents the best-fit redshift ($z=4.19$) from the optical-to-NIR SED.
}
\label{fig:chi2fit}
\end{figure}

We use the available photometry in Table 1 at 24\ums, 100\ums, 160\ums, 250\ums, 350\ums, 500\ums, 850\ums, 1.1 mm and 20 cm (Fig.~\ref{fig:chi2fit}),  
and compare them to a suite of 105 template SEDs from the \citet[][]{ce01} library (CE01 hereafter), as a function of redshift. 
The CE01 templates are sorted by IR luminosities from Log $L_{\rm IR}/$\lsol$=8.41$ to Log $L_{\rm IR}/$\lsol$=13.55$, 
with dust temperatures becoming warmer with increasing luminosity. 
We calculate the $\chi^2$ values between the observed flux densities of \src and predictions for  
each SED model of given luminosity as a function of redshift. We choose a redshift grid from $z=0.1$ to $z=8$ in steps of $\delta z=0.1$. 
We note that an AGN contributes significantly
to the radio emission, higher than the 
predicted value from any SED template by a factor of more than $20$. Such radio excess could bias the best-fit redshift. We are therefore excluding the radio data in our FIR SED fits.    
Figure~\ref{fig:chi2fit} (bottom) shows the $\chi^2$ values for 
each of best-fit SED model as a function of 
redshift (thin curve).
The best-fitting redshift is $z_{\rm phot} = 3.1$ and the 90\% confidence interval is $2.51 < z < 4.35$, 
which agrees with the NIR photometric redshift within the errors (Sect. 3.1). 
We note that this analysis allows us to reject 
the possibility of a highly reddened low-redshift galaxy (i.e., at $z\sim0.6$). 


As mentioned above, the dust temperature can bias the redshift estimation from the FIR to radio SED fittings. 
For instance, SED templates with lower dust temperature can decrease the best-fit redshift. 
This can be seen in Fig.~\ref{fig:chi2fit} (bottom) which shows the best-fit FIR luminosity 
that gives the least $\chi^2$ value at each redshift. Since the CE01 templates are sorted such that 
the dust temperatures become warmer with increasing luminosity, the lower FIR luminosity inferred at $z<2$ 
is suggestive of lower dust temperature for the best-fit template.   
For comparison, we also estimated the redshift of \src by fitting the well-studied SEDs of local galaxy 
Mrk 231 ({an obscured AGN} with warm dust), Arp 220 (luminous starburst with cool dust), and M82 (low-luminosity starburst with warm dust). 
The 24\um flux is excluded in the SED fittings, as it can bias the fit statistics.
The best-fit comes from Mrk 231 at $z=4.9$, 
whilst the fit with the Arp 220 template yields a relatively low redshift of $z=3.7$. 
This is perhaps not too surprising since the Arp 220 SED has a colder dust temperature.       
The best fit of the CE01 templates yielding a lower redshift at $z=3.1$ is perhaps due to the very 
wide range of SED models. Generally speaking, the CE01 models provide good fits in the $z\sim2.5-4.5$ range 
at a 90\% confidence level. 
%
Our analysis suggests that the result of the FIR redshift for \src highly depends on its dust temperature which 
cannot be determined independently.  
{In addition, as shown in Fig. 4 (upper panel), the best-fit
  CE01 SED at z=3.1 overpredicts the observed fluxes at $<10$\um. 
This may be due to the poor calibrations of the optical/NIR part of the
CE01 templates, and/or heterogeneous dust extinction from a more
complicated dust distribution (e.g., Daddi et al. 2009a).
}
Therefore, we favor the NIR redshift of $z=4.19$ for \src in this paper, 
though we cannot fully rule out a lower redshift of $z\sim3.1$ from the
FIR SED fittings. 
We note that \citet[][]{liu18} obtained a redshift of $z=3.988$  by
independent FIR SED fittings in which the \citet[][]{magdis12} library of
main sequence and starburst galaxies was used as a template. 
\begin{figure*}
\centering{
\includegraphics[scale=0.88, angle=0]{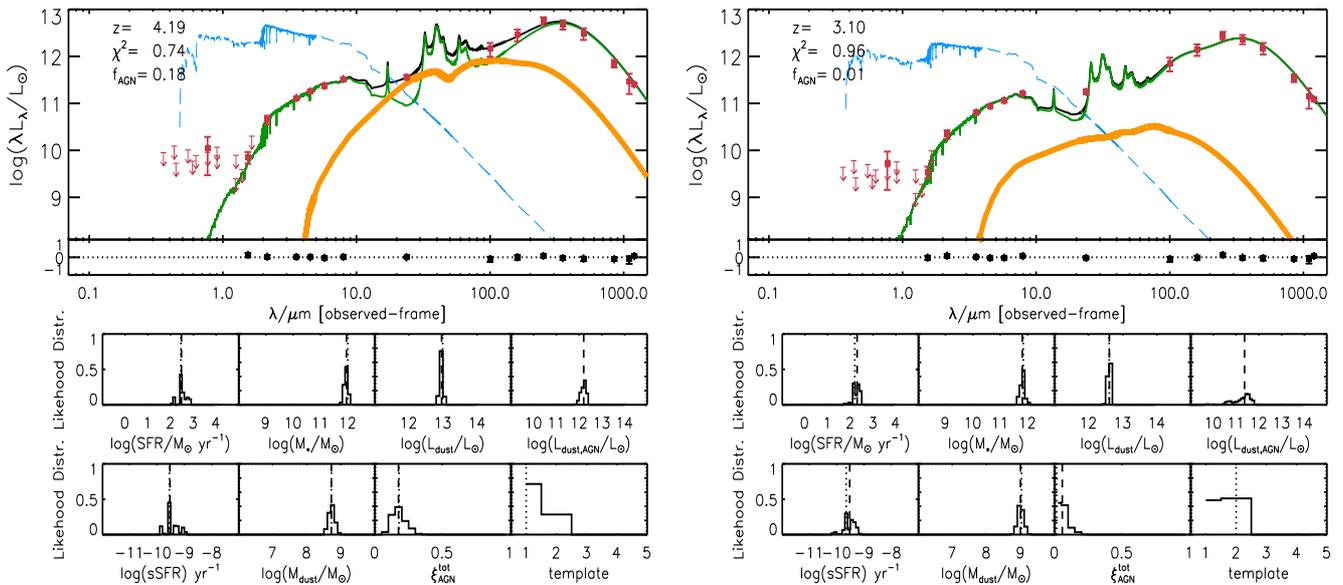}
\caption{
SED fittings to the broadband UV to submillimeter data with {\tt MAGPHYS} at $z=4.19$ (left) and $z=3.1$ (right).
The thick orange line represents the AGN component, with contributions that are estimated at $18$\% and 1\% 
to the total IR SED, respectively.  
The red solid circles are the photometric data and the black lines are
the best-fitting model.  
{The corresponding data to model ratios are shown in the middle panel}.  
The blue line represents the (unattenuated) star formation component. 
The attenuated stellar emission and the reproduced dust emission are 
shown with the green line. 
The posterior probability distribution of best-fitting parameters 
are shown in the lower panel. 
The dashed lines are the median
values and the dotted lines are the best-fitting values.
}\label{fig:magphys}}
\end{figure*}

\subsection{Ultraviolet to far-infrared broadband SED fittings}
It should be noted that the NIR and FIR SED analyses presented above 
were modeling the stellar and dust emission properties 
separately, and do not take into account the AGN component. 
Now, we employ the {\tt MAGPHYS} code \citep[][]{dacunha08, dacunha15} 
to model the UV-optical to submillimeter broadband SED in a more consistent
way. 
{\tt MAGPHYS} is built on a global energy balance between 
stellar and dust emissions: 
the UV/optical emission from stellar continuum 
is absorbed by dust grains in star-forming regions, 
which re-radiate the absorbed energy in the MIR to FIR wavelengths. 
Here we used the latest version of {\tt MAGPHYS} that includes
the contribution by AGN emission to the SEDs (de Cunha et al. in
prep). 
This allows us to disentangle the AGN emission from the broadband SED. 
The AGN emission is modeled using a set of empirical templates from 
\citet[][]{richards06}, \citet[][]{polletta07}, \citet[][]{prieto10}, and \citet[][]{mullaney11},  
which span a wide range of representative AGN SEDs including 
type 2 AGNs, type 1 AGNs and quasars as shown in \citet[][]{chang17}. 
The contribution of the AGN template to the total
IR luminosity is defined as $\xi_{\rm AGN} = L^{\rm AGN}
_{\rm dust} / (L^{\rm AGN}_{\rm dust} +L^{\rm SF}_{\rm dust})$, 
which is allowed to vary between 0 and 1 for each of the templates.  

Since {\tt MAGPHYS} was designed to perform SED fittings with known redshifts, 
we fixed the redshift at $z_{\rm phot}=4.19$ for \srcs. 
The result of the {\tt MAGPHYS} SED fitting is shown in Fig.~\ref{fig:magphys} (left panel), 
which reproduces the broadband SED of \src well, with $\chi^2_{\rm AGN}=0.74$.  
The SED fitting appears to tightly constrain the AGN fraction based on the detection 
of MIR photometric point at the 24\um (which differentiates between AGNs and
star-forming galaxies). 
The posterior probability distribution suggests that
the AGN component contributes at most $\sim$30\% to the total IR luminosity, with 
 $\sim$18\% being the most probable level of contribution. 
The SED fitting assuming the source at $z_{phot}=3.1$ is also plotted 
in Fig.~\ref{fig:magphys} (right panel), which yields an even
smaller best-fitted AGN fraction of $\sim$1\%.  
The dust and stellar properties inferred from the {\tt MAGPHYS}
fittings will be discussed in the following section. 



\section{Properties of \src}
\subsection{A heavily obscured AGN in \src}

\begin{figure}
\centering
 \includegraphics[scale=0.5, angle=0.]{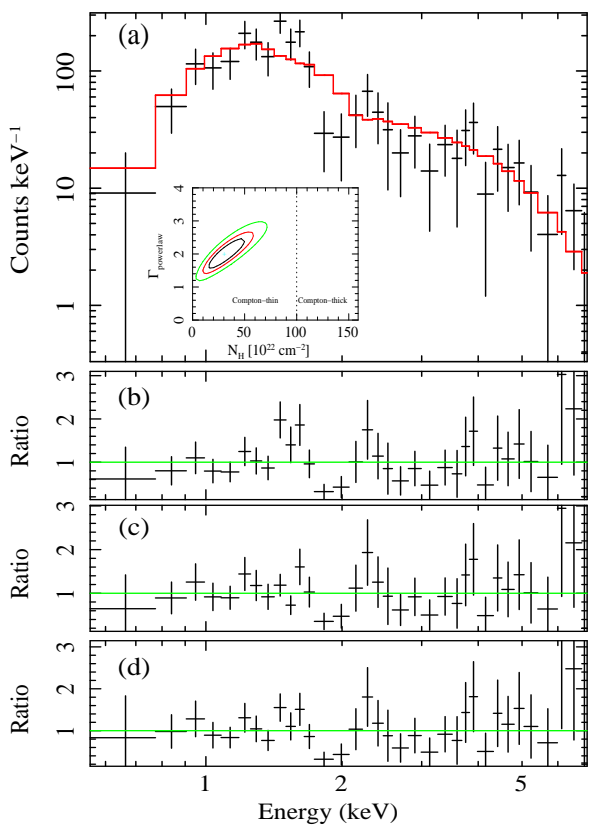}
 \caption{
\chandra X-ray spectrum for \srcs. The red solid line is the best-fit model ({\it wabs*zwabs*zpowerlaw} in XSPEC), 
with the data to model ratio shown in panel (b). 
The inset panel shows the $\chi^2$ contours of the photon index $\Gamma$ vs. column density $N_{\rm H}$ at
the 68\%, 90\% and 99\% confidence levels (from the inside out). 
Panel (c) shows the data/model residuals after including a Gaussian line at $E_{\rm obs}=1.52 keV$. 
Panel (d) shows the data/model residuals by fixing redshift at $z=3.1$. 
}
\label{fig:chandraspec}
\end{figure}

Here we investigate the detailed properties of the X-ray source presented in Sect.~\ref{sec:data}.
Although some results from X-ray spectral fitting have been shown by \citet[][]{laird10}, 
they assumed a much lower redshift of $z=2.6$ for the source, which may result in different 
spectral parameters from ours. 
We emphasize that the analysis presented in this paper is more robust, 
because it is based on 
new constraints on the redshift and updated X-ray data reduction.  
The spectral data were grouped to have at least 1 count per bin and 
the $C$-statistics was adopted for minimization.
Spectral fittings were performed in the 0.5--7 keV band using the \texttt{XSPEC} (Version 12.6). 
 All statistical errors provided hereafter correspond to 90\% confidence for one interesting
parameter ($\Delta\chi^2$= 2.706). 
{ Throughout} the model fittings,
the Galactic column density was considered and
fixed at $N_{\rm H}^{\rm Gal}=0.98\times10^{20}$ cm$^{-2}$ {
  (Kalberla et al. 2005)}.
We performed the spectral fitting with a simple absorbed power-law ($wabs*(zwabs*zpowerlaw)$), assuming $z=4.19$. 
The best-fitting parameters are $\Gamma=1.99^{+0.50}_{-0.47}$ and $N_H=3.3^{+2.0}_{-1.7}\times10^{23}$ cm$^{-2}$, 
with $C=$162.5 for 163 degrees of freedom. 
Such a $\Gamma$ is typical in X-ray bright Seyfert galaxies \citep[e.g.,][]{tozzi06}. 
The measured $N_H$ indicates that the source is heavily absorbed but not Compton-thick. 
The 99\% confidence contour of the $N_{\rm H}$ vs. $\Gamma$ shown in Fig.~\ref{fig:chandraspec} 
implies an upper limit of $N_{\rm H}<10^{24}$ cm$^{-2}$.  
No prominent iron K$\alpha$ line is observed at 1.233 keV (i.e., rest-frame 6.4 keV based on the photo-z).  
We therefore obtained an upper limit on the rest-frame equivalent width of the line of $\sim$760 eV. 
This best-fitting model gives the observed X-ray fluxes of $f_{\rm 0.5-2~keV} = 4.8\times 10^{-16}$ \ergs
and $f_{\rm 2-10~keV} = 1.5\times 10^{-15}$ \ergs.

{ 
As can be see in Fig.~\ref{fig:chandraspec} (a) and (b), 
a positive excess in the fit residuals was present around 1.5 keV. 
Therefore, we added a redshifted Gaussian emission component to test whether or not it is significant. 
The addition of this component improves the fit moderately with
$\Delta\chi^2=8.7$ for two extra free parameters (Fig.~\ref{fig:chandraspec} (c)).
If assuming it to be the Fe K$\alpha$ line at 6.4 keV, the best-fit yields a redshift of 
$z=3.21^{+0.26}_{-0.19}$, and a rest-frame equivalent width of the line $EW_{\rm rest}=603^{+505}_{-419}$ eV.  
The significance of the line detection was at $<$99\% confidence level 
for two parameters (i.e., redshift and normalization). 
If it is real, the best-fit value of redshift is consistent with that obtained from the FIR SED fittings 
($z=3.1$, Sect. 3.2).  
However, due to the limited spectral resolution and statistics of the \chandra data, the possibility for the 
presence of a line emission at $E_{\rm obs}\sim1.52$ keV is still low.     
}
If we fix the redshift at $z=3.1$ in the spectral fittings without a
Gaussian line, the residuals vanish as well (Fig.~\ref{fig:chandraspec} (d)), suggesting that the line feature is likely not real. 
 


The absorption-corrected luminosity in the rest frame 2--10 keV, which can be estimated as 
$L_{\rm X}=4\pi d_{\rm L}^2F_{\rm X}^{\rm C}(1+z)^{\Gamma-2}$$\sim2.6\times10^{44}$ \erg and $1.6\times10^{44}$ \erg , for a source at $z=4.19$ and $z=3.1$, respectively,  
places \src at the low end of the X-ray luminosity range for type-2 quasars.  
Therefore it is likely that \src hosts a heavily obscured AGN (but not Compton thick), 
which is one of the X-ray luminous objects found among the SMG population 
\citep[][]{alexander05, wang13}. 
However, we note that the uncertainties in redshift and the geometry of the obscuring
material might affect the derivation of the intrinsic luminosity.

\begin{figure}
\centering
\includegraphics[scale=0.55,angle=0]{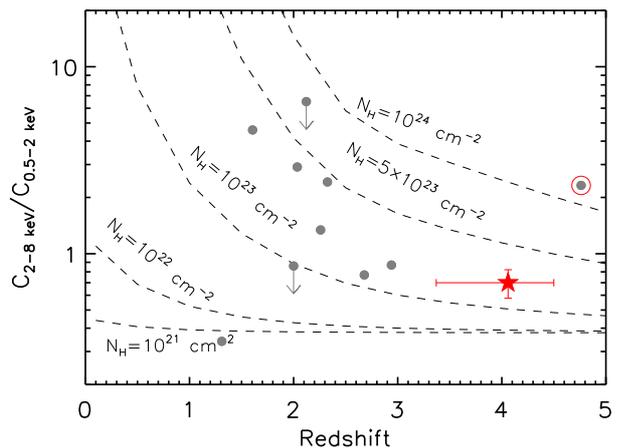}
 \caption{
Hardness ratio between the 2--8 keV (hard) and 0.5--2 keV (soft)
bands (i.e., X-ray hardness) as a function of redshift. 
The dashed lines are simulated tracks for an absorbed power-law with 
photon index fixed at 1.8. 
The red star is the X-ray hardness for \srcs. 
For comparison, we also plot in gray dots the hardness ratios for SMGs in Wang et al. (2013), 
which include the most distant Compton-thick {AGN}, LESS J033229.4-275619 at $z=4.76$ 
(red circle).  
}
\label{fig:simnh}
\end{figure}

We investigated whether the uncertainty in redshift could affect the estimation of the 
amount of absorption from spectral fittings. We run simulations in XSPEC using the 
$wabs*(zwabs*zpow)$ model by fixing $\Gamma=1.8$ and varying $N_H$ to construct hardness ratio 
tracks as a function of redshift (e.g., Wang et al. 2013). 
The results are shown in Fig.~\ref{fig:simnh}. The $N_H$ value constrained in this way is not 
distinguished from the one derived from spectral fitting, suggesting that the \src is 
heavily obscured, with $N_H>10^{23}$ cm$^{-2}$ if at $z>3$. We note that the intrinsic 
luminosity could be considerably greater than the observed luminosity if the obscuration is 
much higher. 




\subsection{Infrared luminosity and star formation rate} 
Given the large X-ray luminosity, the bolometric output from
the central AGN in \src can heat up the interstellar
dust and this may contribute to a substantial fraction of the
FIR luminosity of the host galaxy. 
We derived the FIR luminosity of \src by fitting the data at wavelengths 
greater than 24\um with a suit of CE01 templates assuming the galaxy is located at $z=4.19$, 
and integrating the best-fit SED between rest-frame 8\um and 1000\ums. 
The inferred total IR luminosity is $8.4\times10^{12}$\lsun. Using the SEDs of local galaxies (M82, 
Arp 220 and Mrk 231) as templates yields consistent luminosities in
the range (7.8-8.6)$\times10^{12}$\lsun. 
The unabsorbed X-ray to FIR luminosity ratio of \src is $L_{\rm X}/L_{\rm FIR} = 0.008$. 
This value is very similar to that found by \citet[][]{alexander05} (median $L_{\rm X}/L_{\rm FIR} = 0.004$),
but is an order of magnitude smaller than those found in quasars ($L_{\rm X}/L_{\rm FIR} \sim 0.05$; Elvis et al. 1994; Alexander et al. 2005). The reason why the luminosity ratio is so much smaller than those in quasars is the extreme
FIR luminosity of the source. 
This suggests that the large FIR luminosity of \src is not mainly
powered by the AGN, but should be attributed to intense star formation activities.  
This is consistent with the estimate that {the AGN contributes} at most $\sim$30\% 
of the total IR luminosity from the {\tt MAGPHYS} SED fittings (Sect. 3.3). 
The SFR for \src can be estimated
with the relation $SFR= 1.7 \times 10^{-10} L_{IR}/$\lsun~\citep[e.g.,][]{kennicutt98}. 
Subtracting the 30\% AGN contribution from the total IR luminosity, we obtained 
a SFR of 1001 \sfr~with a Salpeter IMF
{and 623 \sfr~if assuming a Chabrier IMF}. 
We note that  {\tt MAGPHYS}  gives similarly high dust luminosity ($8.0\times10^{12}$\lsun) produced
by the star formation activities. 
This suggests the source to be among the most extreme star-forming galaxies seen
in the early universe (e.g., Casey et al. 2014 and references
therein).

  \begin{figure}
\centering
 \includegraphics[scale=0.3,angle=0]{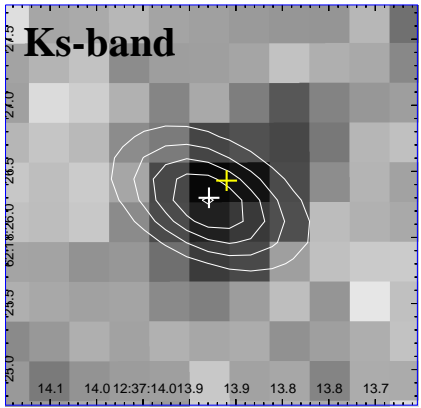}
 \caption{
NOEMA contours (white) overlaid on the $K_s$-band image.  
The white and yellow cross represents the NOEMA 1.2 mm {\bf (beam size of $\sim$1.06\arcsec$\times$0.63\arcsec)}
and $K_s$-band centroid, respectively. The image has a size of
3\arcsec$\times$3\arcsec. 
}
\label{fig:noema}
\end{figure}

\subsection{Stellar population}

As shown in Fig.~\ref{fig:nirsed}, 
the NIR SED of \src is nearly a power law in the IRAC bands but shows a clear spectral break 
between 1.6 and 3.6 \ums. This cannot 
be fully explained by a normal extinction curve. 
The best-fit template from the BayeSED modeling is at $z=4.19$ with the Calzetti extinction of $A_{V}=2.89$. 
In this scenario, the red $H$-[4.5\ums] color is mainly due to the redshifted 4000\AA~Balmer break at 
$z>4$,  which is a signature of existence of older stellar
population. 
The stellar population model from BayeSED  agrees with this interpretation: the best-fit template 
has an age of 0.67 Gyr (maximum likelihood value), which is about half the age of the universe at $z=4.19$.
On the other hand, \src has a very strong FIR emission, very similar to GN10 \citep[][]{wang09}, 
implying an intensive ongoing starburst and the formation of a young galaxy.  
There seems to be a two-component SED model to reconcile the old stellar population 
and very dusty star-forming component seen in the FIR, but the
current data are insufficient for further decomposing the two components in \srcs.  
It is possible that the star-forming component is extremely dusty, such that its 
rest-frame UV and optical light being largely obscured along our line of sight, and 
the NIR SED in the IRAC bands is dominated by an older stellar population with a moderate extinction of 
$A_V<3$.  
We note that the optical/NIR and IRAC data cannot constrain the stellar age well
because of its degeneracy with the dust attenuation. 
By marginalizing a wider range of SFRs, the {\tt
  MAGPHYS} SED fittings yield a stellar age of $\sim0.4$ Gyr with a higher $A_V=4.1$. The galaxy
\src is not resolved by the NOEMA (beam size of $\sim$1.06\arcsec$\times$0.63\arcsec). 
We overlaid the NOEMA contours on the NIR $K_s$-band image, and 
found an offset of $\sim$0.2\arcsec between the NOEMA and $K_s$-band
centroid.  
The latter image is also found to show a clear extension to the northwest (Fig.~\ref{fig:noema}). 
This suggests that the dusty star forming region may be different from that producing  MIR emission. 
{However, though the NOEMA's positional accuracy is high
  ($\sim$0.03\arcsec), the astrometric accuracy of observations at
  $K_s$-band is not as good. 
Wang et al. (2010) have shown that typical astrometric uncertainty at
Ks-band for a source with S/N$>$20 and FWHM$<$1.2" is $\sim$0.05". 
Given the relatively low S/N of \src detected at Ks-band, we estimated a
positional error of $\sim0.2$\arcsec. Such an offset should therefore be considered as tentative. 
}
We note that recent high-resolution submillimeter and/or radio observations have confirmed that some SMGs 
indeed show significant spatial offsets between the positions of dust
emission and the UV/optical 
emission \citep[e.g.,][]{schinnerer08, chen15, hatsukade15, elbaz17}. 


  \begin{figure}[tp]
\centering
 \includegraphics[scale=0.6,angle=0]{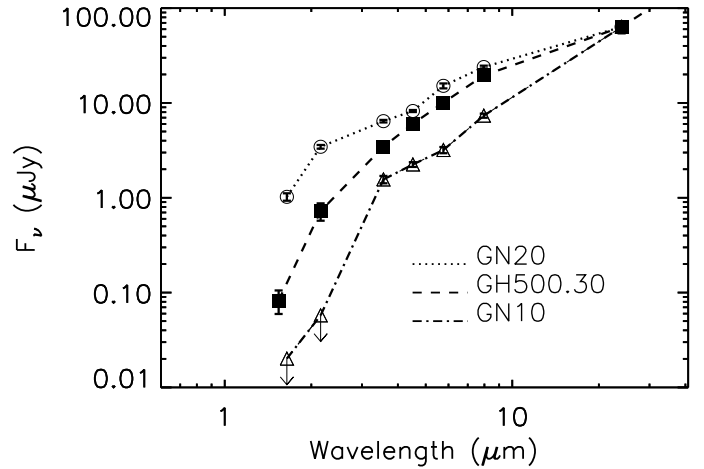}
 \caption{
NIR SED of \src at $\lambda>1$\um (filled squares) for \srcs, 
compared to SEDs of another two bright SMGs at z=4, GN 20 (circles) and
GN 10 (triangles). 
The flux densities are scaled to the 24\um flux of \srcs. 
}
\label{fig:youngsed}
\end{figure}

{We argue that the AGN component is not a major contribution in the MIR. 
Using the intrinsic, absorption-corrected X-ray luminosity from Sect. 4.1, we 
can estimate the rest-frame 6\um luminosity due to the AGN, and compare it with the total 6\um luminosity measured in the MIR. 
Assuming that the {X-ray-to-MIR} luminosity relationship for local AGNs \citep[][]{lutz04} 
holds at high redshift, the intrinsic 2--10 keV luminosity suggests a 6\um 
luminosity of $\nu L_{\nu}(6\mu m)=1^{+2}_{-0.7}\times10^{45}$\erg for the AGN.   
We derived total 6\um luminosity from the 24\um flux (the {MIR} spectral slope of $\alpha=1.8$, 
$S_{\nu} \propto \nu^{- \alpha}$ is used for the small $K$-correction, Coppin et al. 2010), 
and found $\nu L_{\nu}(6\mu m)=8.6\times10^{45}$\erg. 
Taking into account the dispersion of the X-ray to MIR luminosity relation, 
the AGN may contribute  $\sim4-35$\% of the observed 24\um emission, 
consistent with the inferred fraction of AGN contribution to the total
IR luminosity from the {\tt MAGPHYS} fittings (Sect. 3.3).  
As shown in Fig. 5, assuming the 24\um flux is completely due to the AGN, the extrapolation of the
best-fitting AGN SED to the IRAC bands yields a contribution to the IRAC fluxes 
of less than 10\%. 
We note that \citet[][]{riechers14} recently presented a measurement of $\sim30-60$\% AGN 
contribution to the MIR continuum emission in GN20, an X-ray detected bright SMG 
at $z=4.05$, from the {\it Spitzer}/IRS spectral decomposition\footnote{\citet[][]{tan14} have shown that the IRAC fluxes from 4.5\um to 8\um 
exceed the best-fit stellar SED for GN20, 
which is likely due to the contribution by an AGN \citep[][]{riechers14}. 
Such an excess IRAC emission was not found in the best-fitting stellar
SED for \src (Fig. 2), suggesting a relatively weak AGN contribution.}. 
In Fig.~\ref{fig:youngsed}, a comparison between the
NIR SED of GN20 and the \src (scaled to the 24\um flux) is shown. 
This comparison suggests that the presence of an obscured AGN in \src is likely, 
but its contribution to the MIR emission is not as much as in GN 20.   
For comparison, we also plot the SED of the extremely dusty SMG GN 10 \citep[][]{wang09}. 
Because of the strong dust extinction, the AGN contribution to the IRAC flux densities in GN10 
is even smaller, 
which is consistent with its non-detection in the X-rays. 

   
}


Having established that the observed IRAC fluxes are dominated by old stars, we can estimate 
the stellar mass for \srcs. The best-fit SED model from the BC03 stellar population yields 
a stellar mass of $M_\mathrm{star} = 3.5 \times 10^{11} M_{\sun}$.  
We note that the {\tt MAGPHYS} SED fittings yielded an even higher
stellar mass of $8.9 \times 10^{11} M_{\sun}$, which may be due to 
higher dust attenuation $A_{\rm V}$ inferred, as there is a degeneracy
between the two (Fig. 3). 
This mass is similar to those for typical SMGs at $z\simgt4$ (e.g., 
Huang et al. 2014; Micha{\l}owski et al. 2012; Smol\v{c}i\'{c} et al. 2015), but is larger than that of coeval UV-selected star-forming galaxies 
which have stellar masses of $\sim10^{10}-10^{10.5} M_{\sun}$ 
\citep[e.g.,][]{stark07, stark09}.
For comparison, \citet{Borys05} have studied stellar components of 13 SMGs in the GOODS-North field, 
and found stellar masses of $\sim 10^{11}$--$10^{12} M_{\sun}$ and a mean stellar mass of $\sim 2 \times 10^{11} M_\sun$.
Although the posterior probability distribution on $A_{V}$ is weakly constrained (Fig.~\ref{fig:contour}), 
 the maximum likelihood value of $A_{V}$ from the best-fitting model is also comparable to those found 
in SMGs with rest-frame optical 
spectroscopy observations by \citet[][$A_V = 3.0\pm 1.0$]{Swinbank04}, \citet[][$A_V = 1.7\pm 0.2$]{Borys05}, 
and \citet[][$A_V = 2.9\pm 0.5$]{Takata06}. 
To summarize, \src appears to be a massive galaxy with an old stellar
population ($\sim0.4-0.6$ Gyr) and moderate dust extinction of $A_V\sim3-4$, 
which may be spatially offset from the extremely dusty starburst region. 
Given its extreme faintness at wavelengths $\lambda<2$\ums, 
massive galaxies like \src at $z\simgt4$ may be missed in 
deep optical and even NIR surveys with \hst \citep[][]{wang16}. 
%

\section{Discussion}


The galaxy \src shares many properties with GN 10, one of the bright SMGs in the GOODS-North field 
confirmed at $z=4.05$. Both sources are incredibly faint at all optical to NIR wavelengths at $\lambda\simlt2$\ums, 
with extremely red colors of $H$-[4.5\ums]$\simgt$4, indicating the existence of an older stellar population if found to be due to 
a strong, redshifted 4000\AA~Balmer break. One unique property of \src is its significant detection 
in the X-rays and radio, suggesting the presence of an AGN. The full-band $\sim250$ net counts ensure meaningful X-ray spectral fitting, and hence put strong constrains on the absorption {of $N_H=3.3^{+2.0}_{-1.7}\times10^{23}$ cm$^{-2}$. } 
This is one of the most distant heavily obscured AGNs confirmed by direct X-ray spectral analysis.  
The most distant Compton-thick AGN discovered so far is LESS J033229.4-275619 \citep[][]{gilli11}, which is also 
a submillimeter bright dusty starburst galaxy. However, unlike \srcs, its rest-frame UV/optical light is only 
partially obscured along 
our line of sight, suggesting less dust extinction or a different
dust distribution in the star-forming region. 
In addition, no significant radio emission due to the AGN is detected in LESS J033229.4-275619. 
\subsection{Can \src-like objects fit into the evolutionary sequence?}
The formation and co-evolution of AGNs and stellar populations
are focuses of a large amount of current research. 
It is believed that the starburst and the growth of SMBHs are generally
triggered by major mergers of gas-rich galaxies (see e.g., 
Hopkins et al. 2008, and references therein). Intense starburst
rapidly enshrouds a forming
galaxy in cold dust which emits at submillimeter wavelengths.
This is followed by an increase in fueling of the SMBH, triggering
an AGN and generating X-ray and radio emission.
Finally, AGN feedback affects star formation in the host galaxy by 
removing the dust and gas, leaving a quiescent massive galaxy 
\citep[e.g.,][]{sanders88, hopkins06, narayanan10}.
In this scenario, \src may be in the early transition phase from a starburst SMG  
to an obscured quasar where a radio-mode feedback is in place, 
making it a powerful probe of  evolutionary models.



{
\src appears to be a luminous massive starburst, forming stars at a rate 
of $\sim$1000\sfr, with properties representative of SMGs at $z\sim2-4$.    
The presence of an AGN has been suggested for approximately
25\% of the SMG population at $z\sim2$ 
\citep[][]{alexander05,
  wang13}. 
Assuming that all bright SMGs go through a subsequent AGN phase, 
the inferred AGN fraction may represent a ``duty cycle" for BH accretion. 
 The current total lifetime estimates from modeling SMGs are $\sim100-300$ Myr 
\citep[e.g.,][]{swinbank06}, and thus the AGN-dominated phase caught by the X-ray/radio 
would be occurring during the last $\sim 25-75$ Myr of an SMG lifetime.
These relative lifetimes and duty cycle are consistent with the
merger simulations of \citet[][]{hopkins08}. 
It is therefore possible that the BH accretion in \src is at a transition phase to an obscured 
quasar, or a protoquasar phase
where the BH is growing more rapidly than in typical quasars and is
{still not completely buried in the central dust and gas-rich region 
at the approximately parsec scale, as in the dusty torus} \citep[e.g.,][]{alexander05, kawakatu09, tamura10}. 
Tentative evidence for this has been found in typical SMGs where the 
BH-to-galaxy mass ratio is found to be lower than in $z$$\sim$2 quasars 
\citep[][]{alexander08b, coppin08}. 
One puzzling property of \src is that while strong AGN emission is observed 
in the X-rays {and radio}, it does not contribute significantly to the MIR fluxes (Sect. 4.3). 
This may be because the AGN is attenuated by dust on a  larger scale in the host 
(from starburst) rather than a torus component {which may have not yet fully formed in \srcs}. 



}
 
\begin{figure}
\centering
\includegraphics[scale=0.7,angle=0]{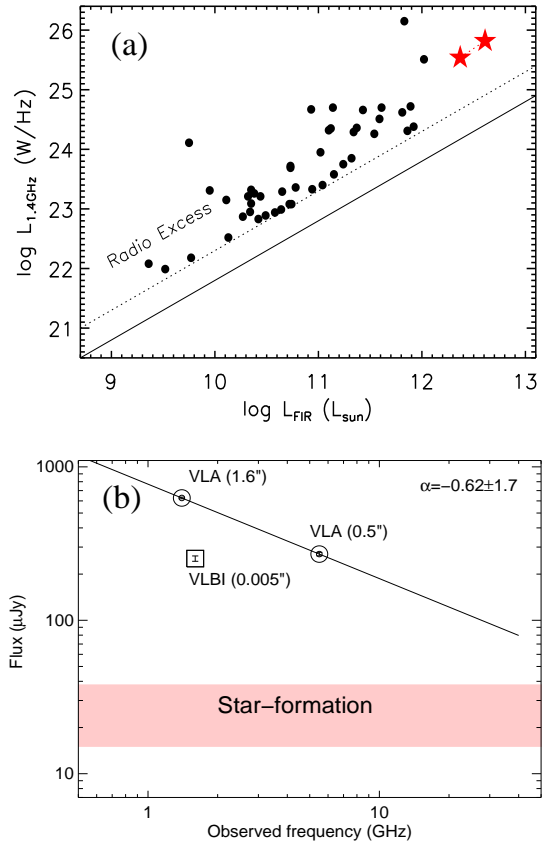}
 \caption{
{Upper panel:} Rest-frame FIR (42.5--122.5\ums) vs. radio 1.4 GHz
luminosity for \src (red star, assuming at z=4.19 and z=3.1, respectively).   
Radio-excess sources from \citet[][]{delmoro13} are plotted as filled circles for comparison. 
The solid line corresponds to $q=$2.21 (average for ``radio-normal"
sources) and the dotted line correspond to $q=$ 1.68, for the
selection limit for radio-excess sources. 
{Lower panel:} Radio slope based on the observations at 1.4 GHz and
5.5 GHz. The 1.6 GHz flux obtained from the high-resolution VLBI observation is shown for
comparison.   
The light-red shaded region shows the 1.4 GHz radio flux expected from the
star-formation process, which is derived by extrapolating the 
best-fitting SED model to 1.4 GHz (Fig. 4). 
}
\label{fig:radioexcess}
\end{figure}

\subsection{Young powerful radio jet in a transitional submillimeter galaxy}

Another unique property of \src is its extremely strong radio emission. 
The rest-frame 1.4 GHz monochromatic luminosity is 6.6$\times10^{25}$ W Hz$^{-1}$, 
and the ratio of rest-frame 1.4 GHz to rest-frame 4400\AA~
(observed $K_s$) flux densities is $\sim$580, which qualifies 
\src as radio loud \citep[][]{stern00}. 
{To derive the rest-frame 1.4 GHz flux, we used a typical radio slope of $S_{\nu}\sim\nu^{-0.7}$ 
(e.g., Thomson et al. 2014) for the K-correction assuming the source at $z=4.19$.}
However, because the observed $K_s$-band flux is dominated by stellar 
population and the AGN optical emission is heavily attenuated,
we should treat this estimation with caution. 
Alternatively, the radio loudness can be measured from the ratio of 
the flux densities between 5 GHz and the X-ray, $R_{\rm X}= \nu L_{\nu}(5GHz)/L_{X}$,
where {5GHz luminosity is in the rest-frame and} $L_{X}$ is the intrinsic luminosity in the 2--10 keV band. We obtained a log$R_{\rm X}$ value of
$-2.29$, placing \src as a radio-loud object where radio-loud is defined as 
 log$R_{\rm X}>-4.5$ \citep[][]{terashima03}.  
%
Using the rest-frame FIR luminosity from the best-fit SED template,  
we obtain $q_{\rm IR}$ = 1.04 which is defined as the logarithmic ratio of 
the rest-frame 8--1000\um flux and the 1.4-GHz flux density. This
is twenty times more radio loud than the normal value of 2.4 $\pm$ 0.24 for FIR-selected 
star-forming galaxies \citep[][]{Ivison10}.
For comparison,
the ratio between the FIR and radio emission in \src is much less than most of the radio excess
sources studied with \herschel in the GOODS-North field
(Fig.~\ref{fig:radioexcess} (a)). 
The detection of a radio-luminous AGN at such a high redshift ($z\simgt4$) is rare, but 
may provide us new insights into the earliest stages of jet formation
and AGN evolution.  

A compact radio core emission has been observed for \src 
with VLBI  at $\sim$5 mas scale \citep[][]{radcliffe16}, corresponding to a
linear size less than 1 kpc at the source redshift. 
This is similar to those of compact steep-spectrum (CSS) radio sources, {a class of AGNs} mostly 
composed of young radio galaxies \citep[][]{odea98}.  
As shown in Fig.~\ref{fig:radioexcess} (b), the VLBI observation
resolves out $\sim$70\% of the radio flux observed with VLA at a resolution of
1.6\arcsec. 
We note that most of the emission that is invisible to high-resolution
VLBA imaging tends to be the extended
component of the jet emission, such as lobes on larger scales, rather
than the star-forming process. 
The latter is estimated to represent only $\sim$5\% of the total radio flux observed by
VLA. Therefore, strong interactions between the jet and the dense
nuclear ISM are anticipated in \srcs, which may signify the earlier
stage of the feedback from AGN jet mechanical energy.  
Based on the VLA observations at 1.4 GHz and 5.5 GHz, we obtained a steep radio
slope of $\alpha\sim-0.62$, consistent with those CSS sources. 
However, such a measurement of radio slope is not robust, because the
resolution is 
different between the two frequencies (by a factor
of $\sim$3). 
Future high-resolution observations with VLBI at other frequencies
will be helpful to better characterize the radio slope, yielding more accurate constraints
on the physical properties of the AGN radio emission, that is, a young
powerful jet in its earliest formation stage.


Theoretically, the formation of jets is thought to be related to the BH and
accretion process \citep[e.g.,][]{celotti01, meier03, ballantyne07}. 
Greene et al. (2006 and references therein) more generally suggested that radio  
loudness (formation of radio jets) is anticorrelated with accretion rate (their Fig. 3) in a sample of
Seyfert galaxies and PG quasars, albeit with large scatter.
This strengthens the similar situation in microquasars (BH X-ray binaries) 
where jets are found at the low accretion state \citep[][]{fender04, fender14}. 
Whether this trend still holds for high-redshift, highly accreting
quasars has yet to be explored. 
Given the possibly high efficiency of BH accretion ($\eta>0.1$), as suggested in SMGs in general,
\src presents a strong case where jet, at least mildly relativistic,
can be produced in high-$\eta$ systems.  
This is important for addressing basic questions as to how jets are formed. 
Although the situation is far from clear, there are some suggestions that 
radio--loud AGNs are associated with rapidly spinning BHs that are ultimately spun-up
by major BH-BH mergers \citep[][]{wilson95, cm11, chiaberge15}. 
This is consistent with the major ``merger" evolutionary scenario for the formation of 
SMGs (like \srcs) that starts with the collision of gas-rich disk galaxies \citep[e.g.,][]{hopkins08},  
seemingly hinting at a possibly important role of BH spin in the making of relativistic jets.  


{It is interesting to compare the properties of \src to the AGN-dominated Hot Dust Obscured 
Galaxies (HotDOGs) which are proposed as tracers of an exceptional rapid and dusty phase 
of quasar/host galaxy co-evolution \citep[e.g.,][]{wu12, piconcelli15, assef15}. 
Recent X-ray observations of a few HotDOGs confirm their harboring of
highly buried AGNs \citep[][]{stern14, piconcelli15}, 
 with absorption column density $N_{\rm H}>10^{23}$ cm$^{-2}$. 
The bolometric luminosities of many HotDOGs exceed $10^{13}$\lsun, 
suggesting them to be among the most luminous objects in the universe, 
which are likely maintained by SMBHs with $M_{\rm BH}\sim10^9$\msun~ 
accreting near the Eddington limit \citep[e.g.,][]{tsai15}.  
The prototype HotDOG, WISE J1814+3412, has shown radio emission exceeding the amount 
expected from the FIR-radio correlation by a factor of approximately ten \citep[][]{eisenhardt12}, 
indicative of a radio-loud AGN. 
Further high-resolution radio observations of HotDOGs show that their radio 
properties are consistent with young radio AGNs \citep[][]{lonsdale15, frey16}. 
If most of the radio emission is indeed due to AGN jets, 
and considering their extreme luminosities and Eddington ratios ($\lambda\simgt1$), 
HotDOGs seem not to fit the simple picture for jet formation \citep[][]{fender04}, as 
we inferred for \srcs. 
However, unlike \srcs, HotDOGs may represent a {late} phase of the AGN-SMG evolution, 
with similar cold-dust components to SMGs \citep[][]{wu12, jones15} but more massive SMBHs.  
}

\section{Conclusions}

Our analyses of multiwavelength observations reveal many unusual properties of \srcs. 
The SED is well constrained at wavelengths $\lambda$$>$2\um thanks to 
the NOEMA interferometer observations, with significant detections by 
\spitzer/IRAC {and MIPS} wavelengths, and by {\it Herschel}, SCUBA2, AzTEC, and VLA from FIR to centimeter wavelengths. 
This galaxy is also detected at the X-ray by \chandra with high significance ($>10\sigma$), allowing for a direct 
spectral analysis.  
However, its flux drops out at shorter wavelengths, displaying a steeply declining spectrum in NIR. 
We analyzed the photometric redshifts of \srcs, using both the NIR and FIR photometric SED fittings, 
yielding a consistent result that the source is likely to be at $z\simgt4$; though a lower redshift 
at $z\simgt3.1$ cannot be fully ruled out. 

Explaining the observed properties of the source requires a two-component SED model with an  
old stellar population dominating the optical-to-NIR SED and a very dusty ongoing starburst {(SFR$\sim$1000\sfr)}  
contributing to the FIR emission that is barely visible in the rest-frame UV/optical. 
The optical-to-NIR SED fitting suggests a modest dust extinction ($A_{\rm V}\sim2.9$ mag) of 
starlight from a stellar component with a mass of $M_\mathrm{star} = 3.5 \times 10^{11} M_{\sun}$ ( 
assuming at $z=4.19$), which may be spatially offset from the starburst region. 
Modeling of the X-ray spectrum of \src suggests that it harbors a heavily obscured AGN with 
$N_H=3.3^{+2.0}_{-1.7}\times10^{23}$ cm$^{-2}$, and a 2-10 keV luminosity of $L_{\rm X}=2.6\times10^{44}$\erg, which is in the range of type 2 quasars. 
The radio emission of the source is extremely bright, which is an order of magnitude more than the 
star-formation powered emission, making it one of the most distant radio-luminous SMGs known so far. 


{In view of the  combination of observed characteristics of the galaxy, we conclude that 
the source appears to have been caught in a rare but critical
early transition stage from a submillimeter galaxy to a dust-obscured quasar}. 
The high redshift of \srcs, if confirmed, will have important implications for the early growth 
of a massive BH and its host galaxy, as well as the formation of jet in the early universe.  
However, its extreme faintness at $\lambda<2$\um hampers any optical and NIR spectroscopic 
observations. 
Blind searches of CO molecular emission line(s) with NOEMA, VLA and LMT are required, 
and may be the only way to spectroscopically identify its redshift. 


\begin{acknowledgements}
This work is supported by the Chinese NSF through grant 11573001 {and 11822301}, 
National Basic Research Program 2015CB857005, the Sino-French LIA-Origin joint exchange program 
and Anhui Provincial NSF (1608085QA06).
Y.Q.X. acknowledges support from the 973 Program (2015CB857004),
NSFC-11473026, NSFC-11421303, 
and the CAS Frontier Science Key Research Program (QYZDJ-SSW-SLH006). 
D.L. acknowledges funding from the 
European Research Council (ERC) under the European
Union’s Horizon 2020 research and innovation programme (grant agreement 
No. 694343). 
Y.H. acknowledges support from NSFC-11773063. 
\end{acknowledgements}

\end{document}